# Force Detection Using a Fiber-Optic Cantilever

Raffi Budakian[a)] and Seth J. Putterman

Physics Department, University of California, Los Angeles, CA 90095

*Abstract:* A force measurement technique has been developed that utilizes a clamped fiber optic element both as a cantilever and as a highly sensitive probe of the static and dynamic displacement of a sample that is mounted near its free end. Light from a 1.5 mW superluminescent diode coupled into the fiber is used to detect displacement with $6 \times 10^{-13} \, \mathrm{m} / \sqrt{\mathrm{Hz}}$ sensitivity for frequencies above 40 kHz. This technique has been used to study the interaction between macroscopic bodies with atomic sensitivity. Here, we report measurements of stiffness of junctions that form when two gold surfaces are brought into contact.

---

[a)] Author to whom correspondence should be addressed; electronic mail: budak@physics.ucla.edu

Scanning force microscopes, of which the atomic force microscope (AFM) is one kind, utilize a stylus, typically made using a silicon cantilever with a sharp tip mounted on one end, to scan and record the interaction of the tip with a surface. The radius of the tip, which can be as small as 50Å, limits the size of the interacting area. In the one and a half decades following the introduction of the AFM by Binnig, Quate and Gerber[1], a variety of techniques have been developed to sense the cantilever deflection with sub-angstrom precision[2]. The most popular of these is optical detection. In an optical deflection measurement, a laser is focused on the cantilever and the reflected light is imaged onto a photodiode[3]. The signal to noise ratio for this method is improved significantly by the ratio of the path length between the cantilever and detector. Another method utilizes optical interference between reflected light from the cantilever and a reference in order to determine the displacement amplitude[4]. Both of these techniques have been shown to have a signal to noise ratio on the order of $10^{-4} \ \text{Å} / \sqrt{\text{Hz}}$.

While the small size of the AFM tip coupled with highly sensitive detection of the cantilever deflection makes it well suited for measuring local atomic interactions[5-11], the miniature size of the cantilever does not permit force measurements between macroscopic bodies. In this letter, we describe a simple technique for force detection that utilizes a segment of a optical fiber as a cantilever to study the static and dynamic interactions between extended objects with atomic sensitivity. This method is similar to shear force feedback detection, which is used in near-field scanning optical microscopes (NSOM) where lateral forces acting on the tip of an oscillating fiber are detected through a shift in the frequency or amplitude of the fiber[12,13]. Our method differs from this technique in two respects: (1) the capability to measure forces through static displacements as well as through frequency shift and (2) the ability to attach macroscopic samples to the end of the fiber. Such a device could prove useful to the study of macroscopic phenomena such as friction, wear, and rupture, which result from collective properties of surfaces in contact. One goal is to use this device to connect the mechanical properties of single conduction channel metal junctions[14] and friction, where many channels are at play.

The instrument, shown schematically in Figure 1a, consists of a single-mode silica fiber optic that is clamped some distance, typically 3-4mm, from the free end. The clamped section of the fiber acts as a cantilever whose stiffness in the transverse (x) and longitudinal (z) directions is isotropic and given by $k_f = 3\pi E_s R^4 / 4l^3$. Here, $E_s$ is the Young's modulus of silica, $R$ is the radius of the fiber-optic and $l$ is the distance between the free end and the point where the fiber is clamped[15]. In order to independently calibrate the cantilever stiffness, the fiber tip is displaced statically and the applied force is measured using a digital microbalance.

The displacement of the free end is measured by coupling a 1.5mW EG&G suplerluminescent laser diode (SLD) into the fiber and placing the exiting cone of light onto a 4-quadrant photodiode. The SLD has a center wavelength $\lambda_c = 850\,nm$ and a $\Delta\lambda = 30\,nm$ spectral width. The device has a small coherence length $l_c = \lambda_c^2 / \Delta\lambda \approx 25\,\mu m$, as compared to a conventional laser diode, which makes it rather insensitive to back reflections and provides a highly stable output. In order to maximize the detection sensitivity, the end of the fiber must be placed an optimum distance $h = h_{max}$ from the active area of the photodiode. The exiting cone of light diverges at an angle $\theta_{max} = N.A.$, where $N.A. = 0.14$ is the numeric aperture of the fiber. The end of the fiber is cleaved flat to ensure that the output has a Gaussian profile given by

$$I_\gamma(x,z) = \frac{I_0}{\pi\sigma^2} e^{-(x^2+z^2)/\sigma^2} \qquad (1)$$

Here, $(x,z)$ are the two components of displacement in the plane of the photodiode, $I_\gamma$ is the normalized intensity in units of $W/m^2$, $I_0$ is the total intensity, and $\sigma \cong h\theta_{max}$. The displacement in the x and z directions is measured by constructing the differential photo currents $\Delta i_x = (i_A + i_D) - (i_B + i_C)$ and $\Delta i_z = (i_A + i_B) - (i_C + i_D)$ (see Figure 1b). The derivative of the measured signal in a given direction with respect to displacement is given by the expression

$$\frac{d\Delta i_n}{dn} = \frac{4I_o\gamma\alpha}{d\sqrt{\pi}} e^{-\alpha^2}\left(1 - erf(\alpha)\right) \qquad (2)$$

Where, $n$ is either the x or z displacement, $\gamma$ is the responsivity of the photodiode expressed in $Amps/Watt$, $d$ is the gap width between photodiode elements, $\alpha = d/2\sigma$ and $erf(\alpha)$ is the error function of $\alpha$. Typically, $\gamma \cong 0.6\,A/W$ for $\lambda = 850nm$. The r.h.s. of (2) is maximum for $\alpha \cong 0.400$, so the highest sensitivity is achieved when $h_{max} \cong 5d/4\theta_{max}$. For the detector used, $d = 13\mu m$, thus, $h_{max} \cong 115\mu m$. The R.M.S. shot noise in a bandwidth $\Delta f$ may be written as $i_{SN}/\sqrt{\Delta f} = \sqrt{2e(1-\kappa)I_o\gamma}$. The fraction of the light blocked by the gap between the photodiode elements is given by $\kappa$ and $e$ is the electron charge. The maximum signal to noise ratio for the displacement signal is

$$\frac{\Delta n}{\sqrt{\Delta f}} \cong \frac{5d}{4} \sqrt{\frac{2\pi e(1-\kappa)}{I_0 \gamma}} \qquad (3)$$

For $h = h_{max}$, $\kappa \cong 0.68$ so that $\Delta n / \sqrt{\Delta f} \cong 3 \times 10^{-13} \, m / \sqrt{Hz}$. Figure 2 shows the R.M.S. spectral noise density taken with a fiber optic cantilever whose stiffness is $k_f = 80 \, N/m$. The peaks at 4 kHz and 35 kHz are the fundamental mode and a higher harmonic of the transverse mode of the fiber optic cantilever driven by thermal fluctuations.

The fiber optic cantilever was designed to study atomic interactions between macroscopic objects. In particular we wanted to measure how bond formation at a metal-metal interface leads to frictional coupling at the macroscopic scale. The interaction between two gold surfaces was measured by gluing a 200μm diameter gold sphere near the end of the clamped fiber. The second surface, a 2mm diameter gold sphere, is attached to a piezo stage used to bring the two surfaces into contact (see Figure 1a). The measurements were carried out after baking the surfaces at 60ºC and $10^{-6}$ torr vacuum for several days and treating all surfaces with 2,2-dichloropropane to drive off residual water[16]. As the surfaces make contact, a gold junction forms at the interface. The stiffness of the junction is an additional restoring force on the cantilever and shifts the modes of the fiber. By measuring the fundamental modes, both the transverse and longitudinal stiffness of the junction can be measured during bond formation. The junction stiffness $k_n$ may be related to the shift in resonance using the expression

$$k_n = k_f \left( \frac{\omega_n^2}{\omega_{n0}^2} - 1 \right) \qquad (4)$$

with $\omega_n = \omega_{n0} + \Delta\omega_{n0}$. The shift in frequency due to the junction stiffness is given by $\Delta\omega_{n0}$ and $\omega_{n0}$ is the free-standing fundamental frequency. During a measurement, the conductance of the junction and the two components of displacement are digitized into a computer as a continuous time record at $f_s = 20\,ks/\sec$ with a 16-bit digitizer having $1.8 \times 10^{-3}\,\text{Å}/bit$ accuracy. The fundamental transverse and longitudinal modes of the fiber are excited by adding a small amount of broadband noise in the range $3.5\,kHz < f_n < 10\,kHz$ to the sample (z) piezo which generates a 0.7Å RMS response at the resonant frequency.

Figure 3 shows the stiffness ($k_L$ and $k_T$) of the junction while pulling the surfaces apart at 0.8Å/s. The shift to the fundamental modes as a function of time is found by breaking up the continuous time record of the displacement into N=200 point sequential segments and Fourier transforming each segment to locate the peak of the resonance. During elongation, the diameter of the junction decreases along with the number of conducting channels as shown in Figure 3a[14,17,18]. The corresponding change in junction stiffness is shown in Figure 3b. It is striking that changes in conductance can display quantized steps of order $10\,G_0$, where $G_0 = 2e^2/h$ is the conductance quantum, even at conductance values as large as 500 units. Agraït[19] and Rubio[18] have previously reported quantized steps at high conductance values for gold junctions. Figure 3 also shows theoretical values for $k_L$ (ç) and $k_T$ (ó) which compare well with the measured data. The two components of stiffness can be expressed in terms of the contact radius $r_c$ using elasticity theory, which gives $k_L = AE_g r_c = 2AE_g \sqrt{n}/k_F$ and $k_T = k_L/2(1+\nu)$.

The Young's modulus and Poisson's ratio of gold are $E_g = 80\,GPa$ and $\nu = 0.42$, respectively. A is a constant, which for these geometries, is order 1. The radius of the junction is determined from the measured conductance using Sharvin's formula[20], $n = G/G_0 = (r_c k_F)^2/4$. The theoretical values shown in Figure 3b are calculated using the conductance curve Figure 3a, taking A=0.60.

The segment of data plotted in Figure 4, which is taken from a different contact-elongation sequence than the data in Figure 3, shows the force applied to the junction as well as its longitudinal stiffness. The stiffness is represented as an image to show the frequency response of the cantilever. The image is constructed by stacking the magnitude spectra of consecutive N=1500 points segments as a function of z-piezo displacement. The frequency resolution is given by $\Delta f = f_s/N$, which for this image is 13Hz/bin. False coloring is used to indicate the response amplitude as a function of frequency. The white points plotted in Figure 4b are calculated using the conductance data shown in Figure 4a with A=0.65. The changes in force and stiffness that accompany the steps in conductance are $\Delta F = 1.0 \pm 0.15\, nN/G_0$ and $\Delta k_L = 0.44 \pm 0.19\, (N/m)/G_0$, respectively. For A=0.65 and n=150, we find $\Delta k_L \simeq 0.35\, (N/m)/G_0$, which is consistent with our measurements.

In conclusion, we have demonstrated a simple and sensitive technique for force detection that is particularly suited for macroscopic objects. The use of a clamped fiber optic as the cantilever affords a great deal of flexibility both for optical detection of displacement and for adjusting the cantilever stiffness. Problems with back reflections, which are common with conventional laser diodes, are avoided by using an SLD that has a very small coherence length and a highly stable output.  This technique has been used to measure the mechanical properties of nano-junctions that spontaneously form at the interface of two macroscopic pieces of gold.

We acknowledge K. Holzcer for valuable discussions regarding the technique and H. Lockart for technical assistance in the construction of the device.  This work is supported by the U.S. Department of Energy (Division of Engineering Research).

# Figure Captions

<u>Figure 1</u> – (a)  Experimental arrangement for measuring bond formation between metal surfaces using the fiber optic cantilever.  The jacket from a segment of the fiber is stripped and the end of the fiber is cleaved flat.  The exposed section is then glued into a glass capillary.  The 200μm gold ball is glued near the free end of the fiber.  Two 10μm diameter gold wires fused to the gold ball are used in the 4-probe conductance measurement of the junction.  The apparatus, which includes the piezo stages and the fiber mount, is suspended from springs and magnetically damped for vibration isolation. (b)  Schematic of the 4-quadrant detector showing the profile of the illuminated region.

<u>Figure 2</u> – R.M.S. spectral noise density of the transverse displacement between 100Hz – 45kHz.  The fundamental mode with the attached gold ball is 4 kHz.  Data is taken with the end of the fiber placed approximately 100μm from the active area of the photodiode. The inset shows a the fundamental transverse (3910 Hz) and longitudinal (4050 Hz) modes excited with thermal noise.  Both modes have Q = 450.

<u>Figure 3</u> – (a)  The conductance of a gold junction, plotted in units of $2e^2/h$, while the junction is stretched at 1Å/s.  The junction is biased with 0.5 VDC and a 5kΩ series resistor is used to limit the current. (b)  Transverse $(k_T)$ and longitudinal $(k_L)$ stiffness of the junction along with the theoretically calculated values.  Data was taken with a 125μm diameter fiber, $k_f = 100\,N/m$.

Figure 4 – (a) Force and conductance measured during elongation at a rate of 0.8 Å/s. The force constant of the cantilever used in the measurement is 80 N/m. Both the conductance and force data have been low pass filtered below 50 Hz. (b) Image showing the longitudinal frequency of the cantilever. The color scale is chosen such that black is zero amplitude and yellow corresponds to $3.7 \times 10^{-2} \, \text{Å}^2/\text{Hz}$. The left axis is the frequency of the fundamental mode and the right axis is the junction stiffness calculated from the frequency shift. The Q of the resonance is in the range 250-350. The noise that is visible in the longitudinal frequency is correlated with the low frequency noise present in the cantilever.

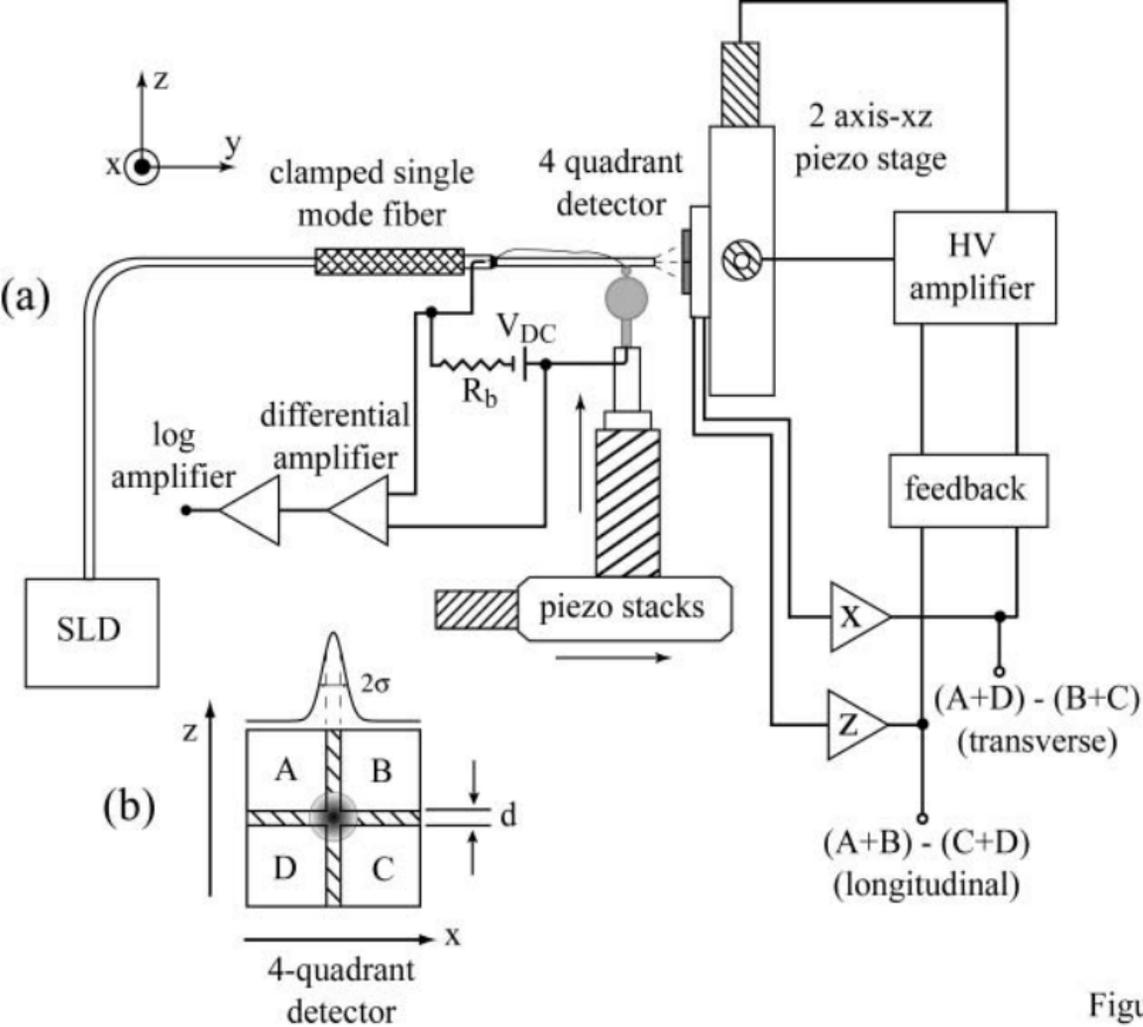

(a)

z
x → y

clamped single mode fiber

4 quadrant detector

2 axis-xz piezo stage

HV amplifier

$V_{DC}$

$R_b$

differential amplifier

log amplifier

SLD

piezo stacks

feedback

X

Z

(A+D) - (B+C) (transverse)

(A+B) - (C+D) (longitudinal)

(b)

z

$2\sigma$

A    B

D    C

d

x

4-quadrant detector

Figure 1

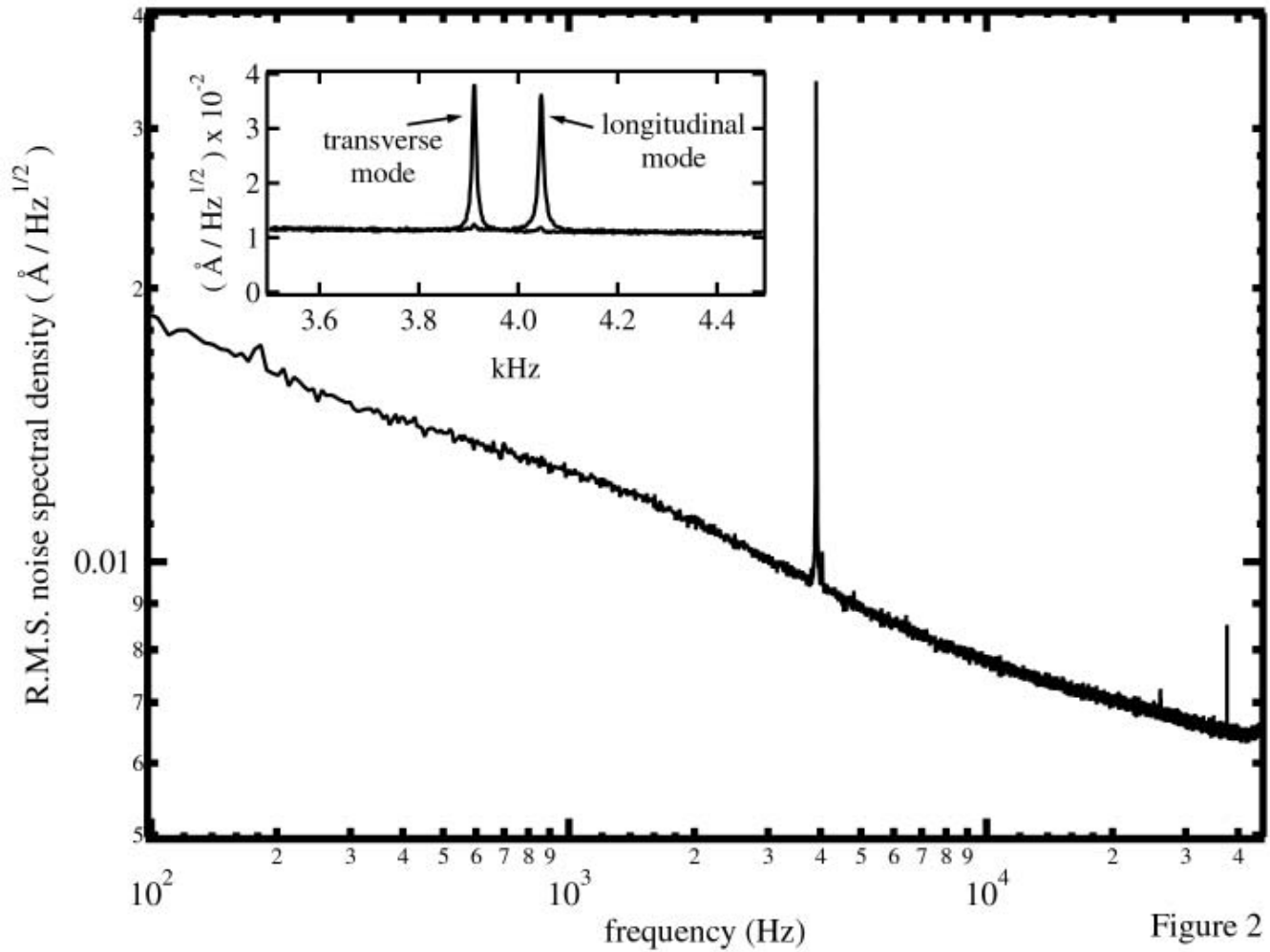

Figure 2

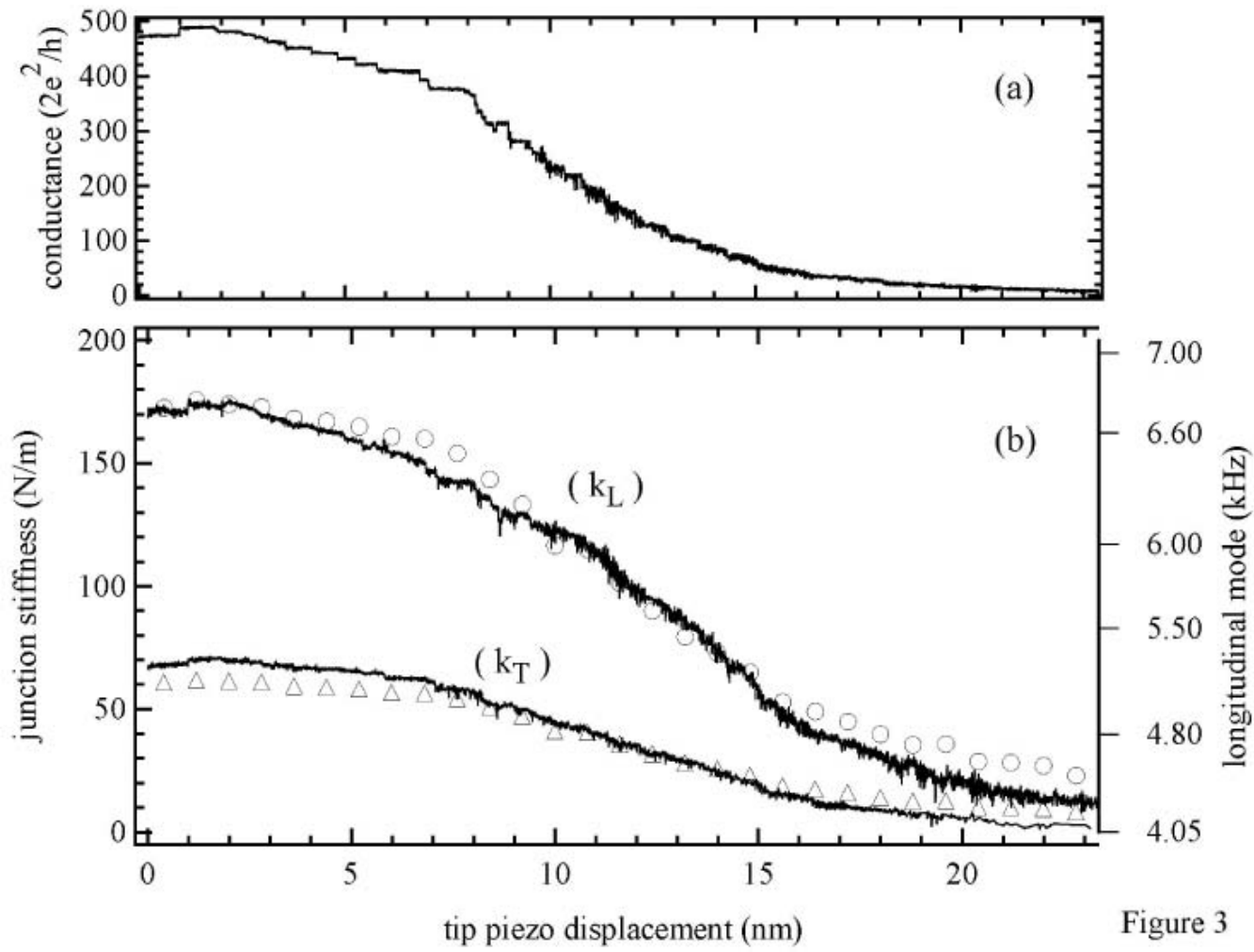

Figure 3

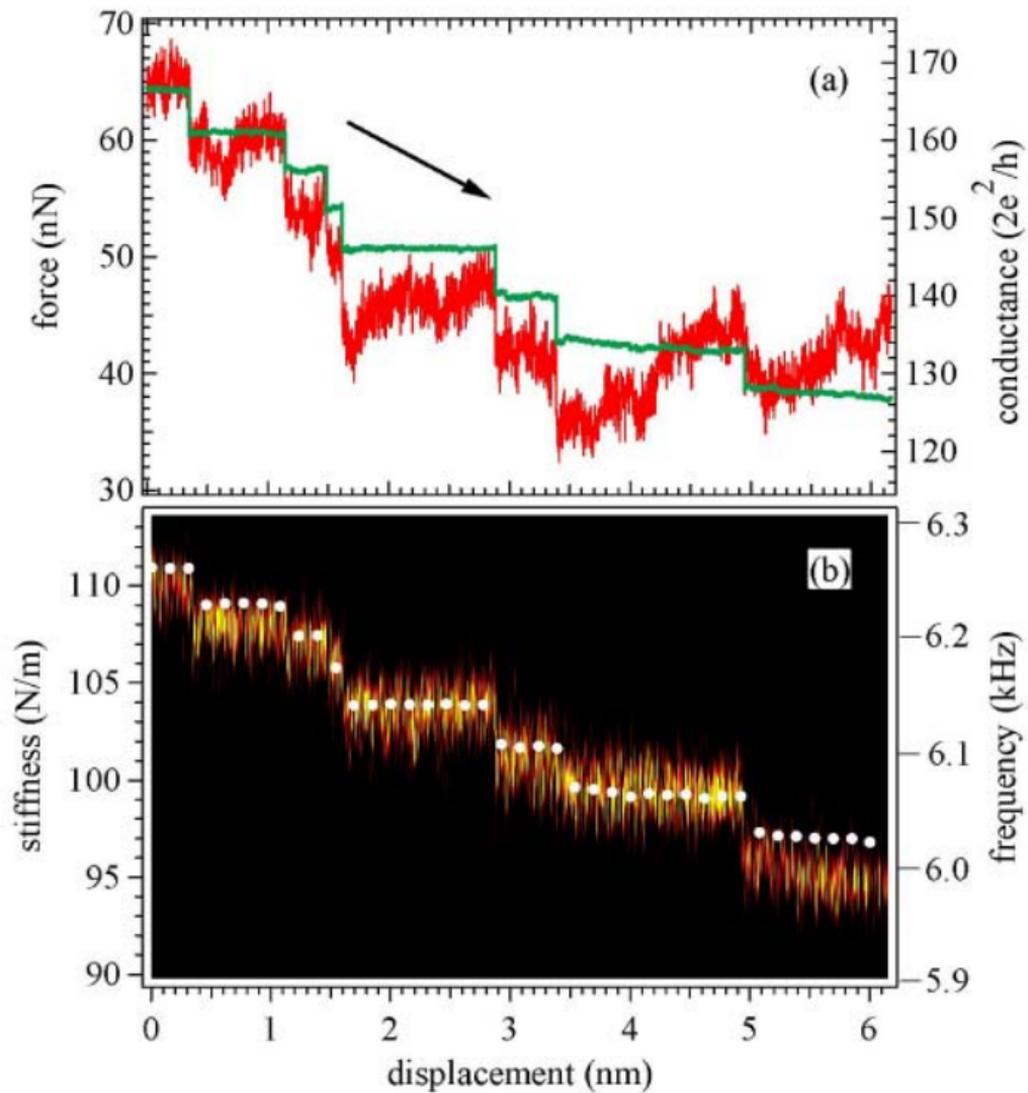

Figure 4